\pdfoutput=1
\documentclass[twocolumn,rmp,superscriptaddress,nofootinbib,floatfix]{revtex4}
\usepackage{dcolumn}
\usepackage{graphicx}
\usepackage{rotating}
\usepackage{multirow}
\usepackage{amsmath,amsfonts,amssymb}
\usepackage{epsfig}

\usepackage{color}

\newcommand{\qed}{\nobreak \ifvmode \relax \else
      \ifdim\lastskip<1.5em \hskip-\lastskip
      \hskip1.5em plus0em minus0.5em \fi \nobreak
      \vrule height0.75em width0.5em depth0.25em\fi}

\begin{document}

\title{The role of gender in scholarly authorship}
\author{Jevin D. West}
\affiliation{Department of Biology, University of Washington, Seattle, WA 98105}
\author{Jennifer Jacquet}
\affiliation{New York University, New York, NY 10012}
\author{Molly M. King}
\affiliation{Department of Sociology, Stanford University, Stanford, CA}
\author{Shelley J. Correll}
\affiliation{Department of Sociology, Stanford University, Stanford, CA}
\author{Carl T. Bergstrom}
\affiliation{Department of Biology, University of Washington, Seattle, WA 98105}\affiliation{Santa Fe Institute, Santa Fe, NM}


\begin{abstract}

Gender disparities appear to be decreasing in academia according to a number of metrics, such as grant funding, hiring, acceptance at scholarly journals, and productivity, and it might be tempting to think that gender inequity will soon be a problem of the past. However, a large-scale analysis based on over eight million papers across the natural sciences, social sciences, and humanities reveals a number of understated and persistent ways in which gender inequities remain. For instance,  even where raw publication counts seem to be equal between genders, close inspection reveals that, in certain fields, men predominate in the prestigious first and last author positions. Moreover, women are significantly underrepresented as authors of single-authored papers. Academics should be aware of the subtle ways that gender disparities can appear in scholarly authorship.


\end{abstract}

\maketitle

%
%

\section{Introduction}

Gender inequities and gender biases persist in higher education.  After decades of high female enrollment in most PhD fields, women represent one-quarter of full professors and earn on average 80\% of the salary of men in comparable positions \cite{WestAndCurtis06}. A recent report \cite{NRC10} surveyed 1800 faculty across six science and engineering disciplines and found men publish significantly more in chemistry and mathematics, while women publish more in electrical engineering (there were no significant differences found in biology, civil engineering, and physics). A recent experiment tested the role of gender in hiring by asking 127 science faculty to evaluate potential lab manager applications and found faculty gave identical applications higher scores if the applicant had a male name \cite{MossRacusinEtAl12}. Another recent analysis of commissioned articles in two prestigious journals published in 2010 and 2011 showed that women scientists are underrepresented; for instance, women wrote just 3.8\% of earth and environmental sciences articles for {\em Nature} News \& Views, although they represent 20\% of the scientists in this discipline \cite{ConleyAndStadmark12}.  With the use of alphabetical authorship listings declining over time \cite{Waltman12}, and given the complexity of evaluating intellectual contributions \cite{Zuckerman68} in increasingly collaborative efforts, understanding patterns of authorship order becomes increasingly important.  

Here we use the JSTOR corpus --- a body of academic papers from a range of scholarly disciplines spanning five centuries --- to examine trends in the gender composition of academic authorship through time. 
We pay particular attention to authorship order, given that first and sometimes last author publications are at least as important as raw publication counts for hiring, promotion, and tenure, particularly in scientific fields \cite{WrenEtAl07}. Studies of authorship in the medical literature reveal, for instance, that women have been historically underrepresented in the prestige positions of first and last author, and that while discrepancies have recently declined in the first author position, women remain underrepresented as last authors \cite{JagsiEtAl06,FeramiscoEtAl09,SidhuEtAl09,Dotson11}.  To view authorship patterns in their disciplinarily context, we use a network-based community detection approach to categorize hierarchically each paper in our study corpus. This yields a hierarchical classification of all papers in our study and allows us to study and compare patterns of gender representation in individual fields of any size and scale.


\section{Methods}

\subsection*{The JSTOR corpus}
The JSTOR corpus (http://www.jstor.org) is a digital archive of published scholarly research that spans the sciences and humanities from 1545 to the present day. At the time of this analysis, the JSTOR corpus  comprised 8.3 million documents ranging from 1545 until early 2011, including 4.2 million research articles. Approximately 1.8 million of these documents (97\% of which are research articles) cite or are cited by other documents in the JSTOR corpus and thus are amenable to network analysis.  We call this group the ``JSTOR network dataset''. Moreover 94\% of these 1.8 million articles are part of a single giant component of the citation network, such that any of these articles can be reached from any other by following citation trails forwards and backwards.   We restrict our analysis to the JSTOR network dataset because this is the portion of the JSTOR corpus that we can hierarchically categorize using citation information. For a list of the main fields available in JSTOR dataset, see Table II.  The gender composition of the identified authors in the network dataset ($21.9\%$ female) is close to that of the identified authors in the entire corpus ($20.8\%$ percent).


\subsection*{Mapping the hierarchical structure of scholarly research}

The scientific literature can be viewed as a large network in which papers are linked by citation relationships \cite{Price65}. The topology of scientific networks can be used to map the structure of science, and the map equation \cite{RosvallAndBergstrom08,RosvallEtAl10} has proven to be a particularly effective method \cite{LancichinettiAndFortunato09}. However, such maps of science have typically shown only a single layer of structure. To map the structure of scholarly disciplines, fields and subfields,  we turn to the hierarchical map equation \cite{RosvallAndBergstrom11}, which reveals multiple levels of substructure within a network. Using the hierarchical map equation on the network of citations, we create a multi-scale map of the JSTOR network dataset in the form of a hierarchical classification that assigns each paper to a major domain, field, subfield, speciality within subfield, and so forth.   For example, Bill Hamillton's classic 1980 paper ``Sex versus asex versus parasite" is classified as residing in Ecology and evolution : Population genetics : Sexual and asexual reproduction : Sex and virulence.  We used the May 13th, 2012 version of the hierarchical map equation code; improvements to that search algorithm made subsequent to our analysis may find somewhat flatter hierarchies than that reported here.  While the algorithm made the decisions about how many fields exist and which papers are assigned to which fields, we manually assigned descriptive names to each field or subfield to facilitate navigation. The names are intended as a general indication of subject matter rather than as a definitive classification.

\subsection*{Determining gender of authors}

We use US Social Security Administration records  to determine gender from first names. The US Social Security Administration website (http://www.ssa.gov/oact/babynames/) makes available the top 1000 names annually for each of the 153 million boys and 143 million girls born from 1880--2010. (These data acknowledge only two genders.)  We assume we can identify an author's gender if the author's first name is associated with a single gender in social security records at least 95\% of the time, as with `Mary', or `John'. Otherwise, as with  `Leslie' or `Sidney', we are unable to identify the gender and do not include that author in our analysis. Since in any given era, androgynous names are more likely to be females, this may slightly downwardly bias our estimates of women \cite{LiebersonEtAl00}.  Similarly, we are unable to classify names that never appear in the top 1000 for either gender in the US records. As a result, authors of some nationalities may be underrepresented in our data set. In a few rare cases national differences may cause misleading assignments for non-US authors (e.g. `Andrea' is typically a female name in the US but a male name in Italy). By this method we are able to assign genders to 6879 unique first names: 3809 female and 3070 male. 

We extracted the first names of all authors in the JSTOR network dataset, discarding those authors who list only initials.  An {\em instance of authorship} consists of a person and a paper for which the person is designated as a co-author. There are 3.6 million authorships in the JSTOR network dataset; of these we are able to extract a full first name for 2.8 million authorships (77\%) associated with 1.5 million papers\footnote{The exclusion of authors with only first initials may exclude women authors disproportionately, particularly in early eras when women may have been more likely than men to publish with initials to avoid potential discrimination.}.  Of these 2.8 million authorships with full first names, we are able to confidently assign gender to 73.3\%. The remaining authorships involve names not in the US social security top 1000 lists (24.3\%), or names associated with both genders (2.4\%).  The final data analyzed include all papers where we know the gender of one or more authors.

 \subsection*{Gender and authorship order}
 
We look at the gender composition of all papers with any number of authors in the JSTOR network dataset.  For every field, subfield, and so-forth, we calculate both the overall gender composition and the gender composition of each authorship position---first, second, third, etc. In some fields, such as molecular biology, the last author position of a paper conveys a special meaning: the last author is typically the principal investigator or group leader of multi-author effort. This is especially the case for papers with at least three authors. Therefore we also report the gender frequency in the last-author position for all papers with three or more co-authors. We then compare the gender frequencies at each author position with the overall gender frequency in the same field. If authorship order were gender-unbiased, we would expect to see the field-wide gender composition reflected at each author position.

\section{Results}

\subsection*{Authorship and author order}

\begin{figure}
  \begin{center}  \includegraphics[scale=.5]{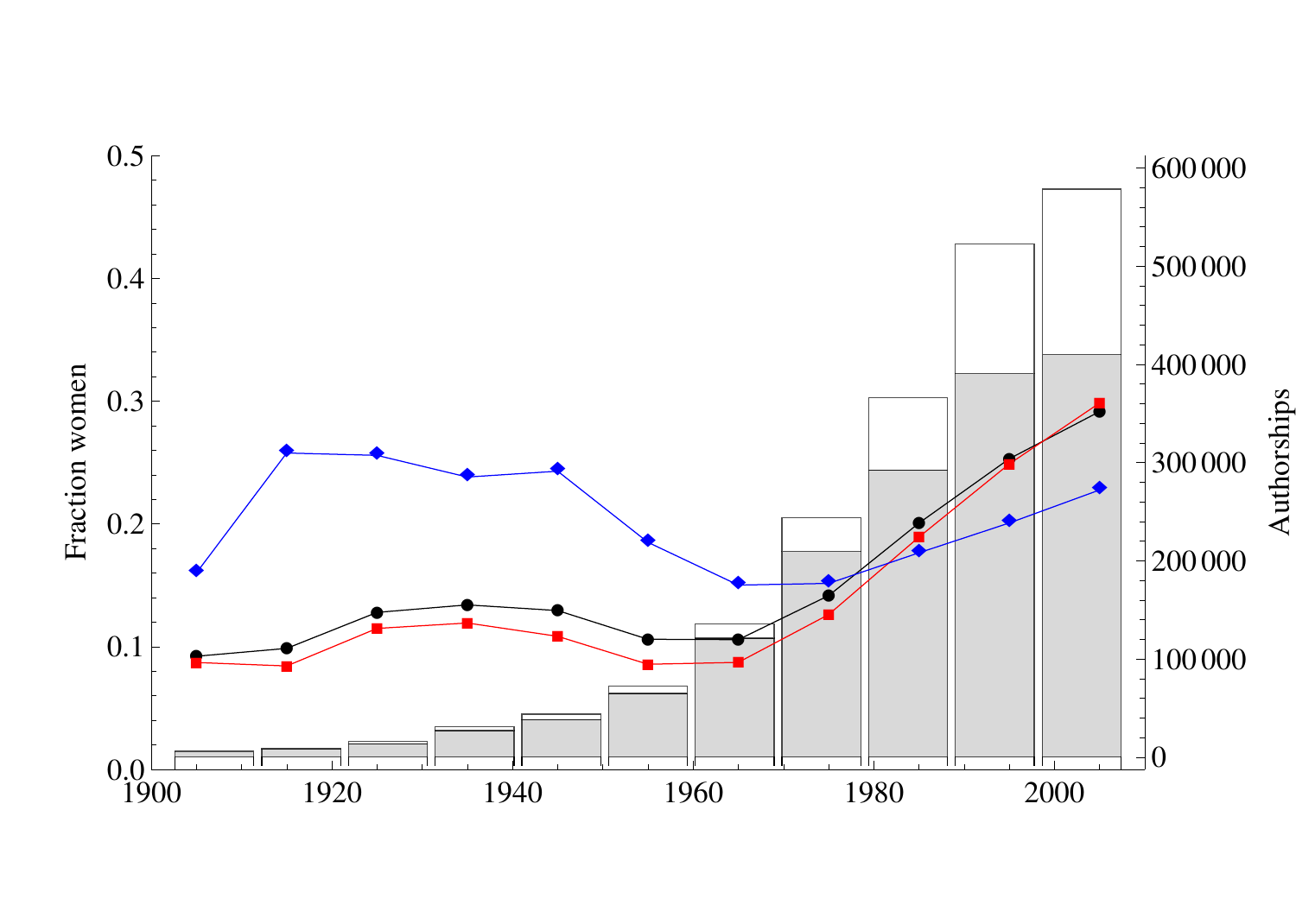} 
  \end{center}
\caption{Authorships and gender composition in the JSTOR network dataset, by decade. Shaded bars represent male authorships, unshaded bars represent female authorships. The black line indicates the fraction of authorships that are women, the red line indicates the fraction of first authorships that are women, and the blue line indicates the fraction of last authorships that are women.  }
\label{fig:genderfreq}
\end{figure}

In an interactive online visualization at {\tt http://www.eigenfactor.org/gender/}, we report the gender composition by authorship position and overall, for each field, subfield, etc., of the JSTOR network dataset.
Women represent 21.9\% of the gender-identified authorships in the entire JSTOR network dataset, but these authorships are not distributed evenly in time across fields, or across authorship positions.  For instance, women represent 17\% of total single-authored papers in the JSTOR network dataset, but represented only 12\% prior to 1990, while they account for 26\% of single-authored papers after 1990. Figure \ref{fig:genderfreq} shows that the fraction of female authorships in general has increased substantially since the 1960s. 
However, some of this increase may result from increased ease of identifying woman authors as individuals become more likely to use first name instead of merely initials.  

\begin{figure}
  \begin{center}
   \includegraphics[scale=0.275]{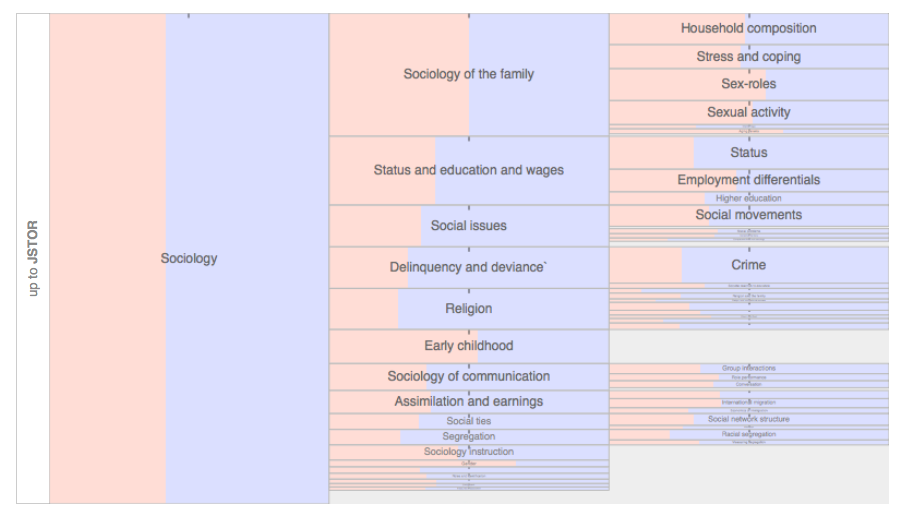}
    \end{center}
\caption{ Even in fields with a gender composition near parity, men (blue bars) and women (pink bars) are unequally distributed in subfields. Shown here is sociology and its subfields from 1990 to the present. An interactive version of this graph, covering all fields and subfields of the JSTOR network dataset, is available online at {\tt http://www.eigenfactor.org/gender/} 
}
\label{fig:uneven}
\end{figure}

Studies of the economics literature have noted considerable differences in gender representation in subfields \cite{BoschiniAndSjogren07,DoladoEtAl05}, and our analysis reveals a comparable pattern across the subfields within the JSTOR network dataset. Even within a field such as sociology that has a relatively even gender balance, different subfields can vary dramatically in gender composition, as illustrated in Figure \ref{fig:uneven}.

\begin{figure}
  \begin{center}
   \includegraphics[scale=0.4]{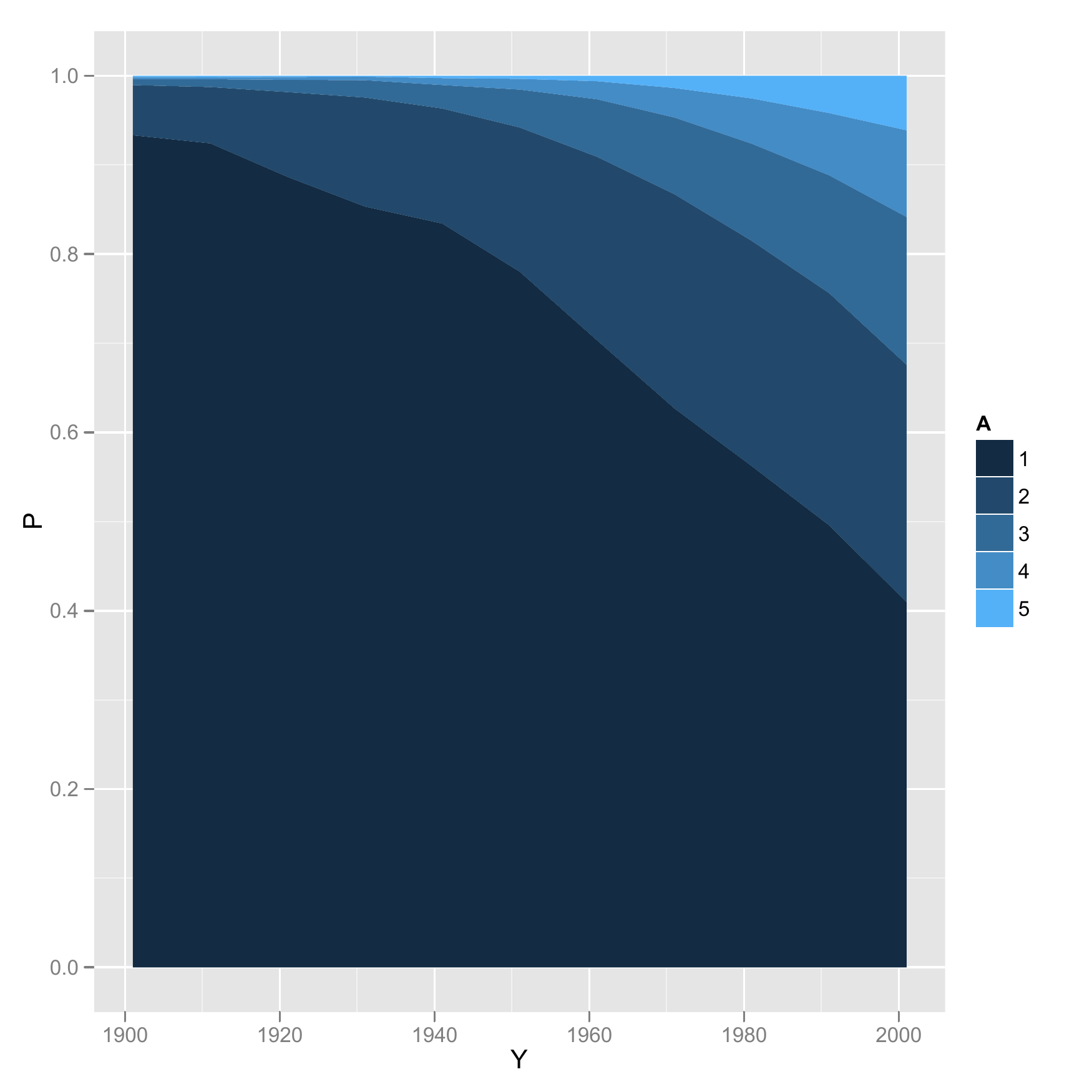}
    \end{center}
\caption{Distribution of author number over time for the JSTOR corpus. Multi-authored papers have increased over time while the fraction of single-authored papers have declined.} 
\label{fig:number_of_authors}
\end{figure}

\begin{figure}
  \begin{center}
  \includegraphics[scale=0.50]{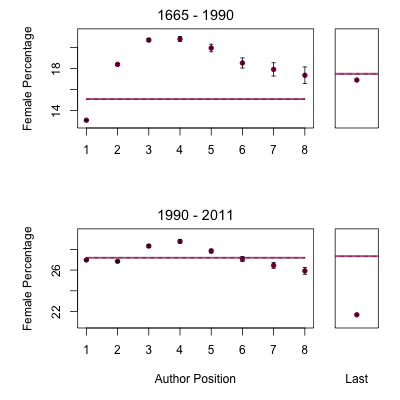}

  \end{center}
\caption{Gender as a function of authorship order across the entire JSTOR network dataset. Top panel: 888,060 authorships prior to 1990. Bottom panel: 1,156,354 authorships from 1990 to the present. From 1990 to present, women are no longer severely underrepresented as first author, but they are increasingly underrepresented as last author.   Error bars indicate one standard deviation of the binomial distribution. For the graph of author position, the solid line indicates the overall frequency of women in the JSTOR network dataset. For the last-author graph, the point indicates the frequency of women who are last author on papers with at least three authors. The horizontal line in this part of the graph indicates the appropriate comparator: the overall frequency of women in any authorship position on papers with three or more authors.  }
 
\label{fig:overall_authorship_order}
\end{figure}

\begin{figure}
  \begin{center}
  
  \includegraphics[scale=0.5]{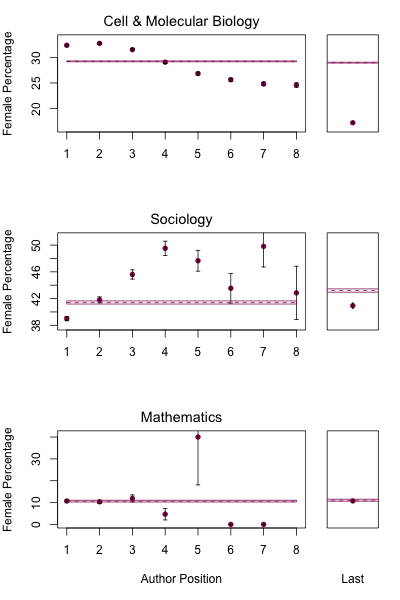}\end{center}

\caption{Gender as a function of authorship position in three domains of scholarship from 1990 to present: cell and molecular biology (276,992 authorships), sociology (44,895 authorships), and mathematics (6,134 authorships). In molecular biology, women are overrepresented as first author but underrepresented at the last author position. In sociology, women are underrepresented in both first and last author positions. In mathematics, where the convention is for alphabetical author order \cite{Waltman12}, women are neither under- nor over-represented at first or last author positions. }

\label{fig:specific_fields}

\end{figure}

Women are not evenly represented across author positions. Prior to 1990, women were significantly underrepresented in the first author position; subsequent to 1990 much of this gap has been closed. However, a new gender gap has emerged in the last author position --- a position of prestige in the biosciences which represent more than half of the authorships in the JSTOR network dataset (Figure \ref{fig:overall_authorship_order}).  Authorship order patterns vary among fields as well (Figure \ref{fig:specific_fields}).  And because conventions of author order vary across disciplines \cite{Endersby96, Waltman12}, underrepresentation of women in the last author position does not hold up in all fields. In mathematics, for instance, author order tends to be alphabetical irrespective of contribution, and in this field women are evenly represented---albeit at low frequency---across authorship positions.

As expected \cite{WuchtyEtAl07}, the proportion of multi-authored papers has increased over time (Figure \ref{fig:number_of_authors}). Some of the pattern in authorship order may be an artifact of this trend in parallel with an increase in the fraction of women over time.

\begin{table}[htdp]\begin{center}
\begin{tabular}{lccccc}
&1960s&1970s&1980s&1990s&2000s \\
\hline
{\bf \% PhDs overall}&8&15&27&33&39\\
Computer sciences&n/a&n/a&12&16&20\\
Engineering&$<$1&2&6&12&18\\
Mathematics&6&11&16&22&28\\
Physical Sciences&4&8&15&22&27\\
Psychology&20&31&49&63&68\\
Social Sciences&10&18&31&38&44\\
{\bf \% Tenure track faculty}&n/a&n/a&13&19&26\\
Full Professors&n/a&5&7&11&18\\
{\bf \% Authors overall}&10.6&14.2&20.1&25.3&29.2\\
\% Single author&8.7&12.5&18.7&24.5&28.5\\
\% 1st author&9.2&12.9&19.3&25.3&30.9\\
\% 2nd author&14.8&16.2&20.8&25.0&28.8\\
\% Last author&15.0&15.2&17.6&20.1&22.8
\end{tabular}
\end{center}
\label{tab:data_table}
\caption{Percentage of women relative to total PhDs and percentage of women in tenure or tenure track positions and full professorships in Science and Engineering from 1960-2006 (source: Burrelli, 2008) as well as percentage of women in various author positions from 1960-2009 as a result of this analysis. 1st and 2nd author positions are listed for papers with at least two authors. Last author percentage is for papers with at least three authors.}
\end{table}
\nocite{Burrelli08}

\begin{table}[htdp]
\begin{center}

\begin{tabular}{lrr}
{\bf Field}&{\bf \% female}&{\bf authorships}\\
\hline
 \text{Mathematics} & 10.64 & 6134 \\
 \text{Philosophy} & 12.04 & 12190 \\
 \text{Economics} & 13.68 & 69142 \\
 \text{Probability and Statistics} & 18.11 & 28324 \\
 \text{Political science - international} & 19.07 & 14908 \\
 \text{Political science-US domestic} & 19.09 & 15705 \\
 \text{Ecology and evolution} & 22.76 & 279012 \\
 \text{Law} & 24.21& 18503 \\
 \text{Organizational and marketing} & 25.44 & 32119 \\
 \text{Physical anthropology} & 27.05 & 16296 \\
 \text{Radiation damage} & 27.69 & 7825 \\
 \text{Classical studies} & 28.88 & 6372 \\
 \text{Molecular $\&$ Cell biology} & 29.25& 277032 \\
 \text{History} & 30.47 & 15585 \\
 \text{Veterinary medicine} & 31.81 & 10960 \\
 \text{Cognitive science} & 32.12 & 12786 \\
 \text{Anthropology} & 36.46 & 19900 \\
 \text{Pollution and occupational health} & 37.57& 32108 \\
 \text{Sociology} & 41.41 & 44895 \\
 \text{Demography} & 41.90 & 7600 \\
 \text{Education} & 46.35 & 28635\\ 
 \label{tab:fields_table}\end{tabular}
\end{center}

\caption{Gender composition from 1990-2011 for disciplines (i.e., groups at the first level of hierarchical clustering) with at least 5,000 authorships.}
\end{table}

\section{Discussion}

Only a century ago, women were forbidden from seeking degrees in most universities in Europe \cite{EtzkowitzEtAl00}.  Women seeking a role in academia faced --- and continue to face --- difficulties at every stage, from admission (Magdalene College at the University of Cambridge was the last all-male college to become mixed, which occurred in 1988), to post-doctoral fellowships \cite{WennerasAndWold97}, to hiring \cite{MossRacusinEtAl12}, to tenure \cite{SpelkeAndGrace06}.  As both women and the belief that they belong in universities have infiltrated the academic system, the situation has greatly improved. Women have earned a higher proportion of bachelor's degrees than men since the mid 1980s \cite{EnglandLi06}.  In 2004, 48\% of PhD recipients were women, up from 16\% in 1972 \cite{WestAndCurtis06}. Despite this increasing equity early in the pipeline, women are still significantly underrepresented in tenure-track and research university faculty positions. Women occupy only 39\% of full-time faculty positions and make up an even lower percentage of full professors \cite{WestAndCurtis06}.  

Since academic publishing is very important to being hired as a faculty member and being promoted, the under-representation of women as authors in academic publications and in more prestigious authorship positions affects the representation of women faculty in academia. Our research shows that women are increasingly represented in JSTOR network dataset authorships: 27.2\% of authorships from 1990-2012 are women compared to just 15.1\% from 1665--1989. However, our results also show that the academic publishing environment remains inequitable.  For instance, since 1990, women represent only 26\% of single-authored papers in the JSTOR dataset.  

In many fields, it is not just sheer number of publications, but author order that matters in promotion and tenure decisions.  Here we show that women have traditionally been underrepresented in the first author position, though this is changing, and women remain underrepresented in the last author position\footnote{Given these findings, we note the irony of our own authorship order on the present paper.}. We should expect some lag between disparity in the first and last author positions, as it takes time for younger scholars to become leaders of research groups. But the difference between total female authorships and first authorships has been less than 2\% since the 1960s, while the discrepancy between total and last authorships remains above 5\%.

While our analysis can clearly delineate gendered patterns in authorship, the data do not allow us to uncover mechanisms that produce the gender disparities we find.  One possibility is that women submit fewer papers than men or that their contributions to papers are less significant than their male coauthorsÕ, thereby landing them in lower prestige positions on papers.  While there is no evidence to support the claim of women's lesser contributions, women are less likely to be involved with collaborative research projects in many scientific fields \cite{Fox01}. A second possibility is that in informal negotiation among a team of authors about author position order, men negotiate more forcefully for the more prestigious positions. While we know of no studies that specifically examine authorship negotiations, men, in general, do negotiate more than women \cite{Babcock07} and are more likely to self-promote their accomplishments \cite{Rudman98}.  A third possibility is that there is a bias against women in the review process, such that when they are in the more prestigious author positions, papers of equal quality are less likely to be accepted than when men occupy the prestigious positions. This would produce an underrepresentation of women in journals that do not rely on gender blind reviews.  While some have claimed, using correlational data, that gender bias is no longer a factor in producing gender disparities in academia  \cite{CeciAndWilliams11}, controlled laboratory experiments and field experiments continue to find that biases negatively affect judgments of women \cite{GoldinEtAl00, CorrellEtAl07}.  For example, a female applicant for science lab manager positions was less likely to be hired than an otherwise identical male applicant, based on judgments of competence by prospective hiring faculty \cite{MossRacusinEtAl12}. Furthermore, the report  ``Beyond Bias and Barriers" reviewed the large literature on gender, bias and academic careers and concluded that subtle biases continue to affect women's careers in academia \cite{NAS07}. 

Our analysis reveals several important patterns: while there have been important gains in parity in the first author position, with the proportion of women in first author positions now even slightly exceeding the overall proportion of female authorships, the proportion of women in the last author position and the proportion authoring overall remain disproportionately low. One strength of this study is that the large dataset represents a significant number of all academics, women and men, across many fields of study and over a large timespan. Though significant progress has been made toward gender equality, important differences in positions of intellectual authorship draw our attention to the subtle ways gender disparities continue to exist. The finding underscores that we cannot yet disregard gender disparity as a notable characteristic of academia.

\newpage

\bibliography{gendauth}

\section*{Acknowledgements}This work was supported in part by NSF grant SBE-0915005 to CTB, NSF Graduate Research Fellowship grant DGE-1147470 to MMK , and a generous gift from JSTOR. We thank Cecilia Aragon and Michael Brooks for their help in designing and implementing the online interactive visualization associated with this paper.

\section*{Author contributions}

JW, JJ, and CB conceived the project. JW wrote the code, compiled the data, and built the interactive browser. All authors were involved in interpreting the data and writing the manuscript. 

\end{document}